# Feasibility Study of Logic Circuits with Spin Wave Bus

[1)]Alexander Khitun, [2)]Dmitri E. Nikonov, [1)]Mingqiang Bao, [1)]Kosmas Galatsis, and [1)]Kang L. Wang

[1)] Device Research Laboratory, Electrical Engineering Department,
MARCO Focus Center on Functional Engineered Nano Architectonics (FENA),
Western Institute of Nanoelectronics (WIN),
University of California at Los Angeles, Los Angeles, California, 90095-1594
[2)] Technology &Manufacturing Group, Intel Corporation, Santa Clara, California, 95054

**Abstract**

We present a feasibility study of logic circuits utilizing spin waves for information transmission and processing. As an alternative approach to the transistor-based architecture, logic circuits with spin wave bus do not use charge as an information carrier. In this work we describe the general concept of logic circuits with spin wave bus and illustrate its performance by numerical simulations based on available experimental data. Theoretical estimates and results of numerical simulations on signal attenuation, signal phase velocity, and the minimum spin wave energy required per bit in the spin bus are obtained. The transport parameters are compared with ones for conventional electronic transmission lines. Spin Wave Bus is not intended to substitute traditional metal interconnects since it has higher signal attenuation and lower signal propagation speed. The potential value of spin wave bus is, however, an interface between electronic circuits and integrated spintronics circuits. The logic circuits with spin wave bus allow us to provide wireless read-in and read-out.

Introduction

There is a growing interest in novel nanometer scale devices and architectures to address the shortcomings and drawbacks inherent to the traditional CMOS-based architecture [1]. Spintronics is one of the most prominent approaches in offering an alternative route to the traditional semiconductor electronics. Recent breakthroughs in the experimental study and control of the spin dynamics in semiconductor nanostructures [2-4] opens new possibilities for spin utilization in information processing. There are some intriguing ideas on possible spin-based logic devices [5-8] taking advantage on the additional degree of freedom provided by spin. It will be extremely beneficial in terms of power consumption to avoid the use of electric current in spin-based circuits. As a possible solution, it was proposed the single spin logic architecture, where the interconnection between the spin-based cells is accomplished via exchange coupling [9]. The use of single electron or single-spin based devices inherently possesses fundamental drawbacks associated with scheme reliability and fabrication tolerance. To overcome these limitations, it was recently proposed the utilization of spin waves as a collective physical phenomenon for information transmission [10].

Spin wave is a collective oscillation of spins in an ordered spin lattice around the direction of magnetization. The phenomenon is similar to the lattice vibration, where atoms oscillate around its equilibrium position. Potentially, it is possible to use ferromagnetic films as spin conduit of wave propagation or referred to as – Spin Wave Bus (SWB), where the information can be coded into a phase of the spin wave. The key advantage of the SWB is that information transmission is accomplished without an electron transport. Besides, there are other significant advantages: (i) ability to use superposition of spin waves in the bus to achieve useful logic functionality; (ii) a number of spin waves with different frequencies can be simultaneously transmitted among a number of spin-based devices; (iii) the interaction between spin waves and the outer devices can be done in a wireless manner, via a magnetic field. The excitation of spin waves can be done by the local magnetic field produced by micro- or nano-scale antenna, while the detection of the spin wave is via the inductive voltage produced by propagating spin waves. The concept of SWB is promising for implementation in different spin-based nano architectures [10]. A first working spin-wave based logic circuit has been recently experimentally demonstrated [11]. The aim of this work is to offer different device principle and to provide a more detailed study of the fundamental limitations and shortcomings of the logic circuits with SWB. One of the most important issues is energy dissipation caused by spin wave damping in the bus, while a low spin wave group velocity is another significant drawback. To understand the transport parameters of the spin bus, we compare them with those for conventional electronic transmission lines. The rest of the paper is organized as follows. In the next section we describe the general principle of the SWB. In section III we consider possible magnetic materials for fabrication and integration of spin waves with conventional silicon platforms. Section IV is devoted to the spin bus transport characteristics, including signal attenuation and signal phase velocity. In Section V we analyze power consumption in SWB; discussions and conclusions are given in sections V and VI, respectively.

II.  The Principle of Operation

The general concept of the logic circuit with Spin Wave Bus is illustrated in Fig.1(a). The input data are received in the form of voltage pulses. For simplicity, we assume that input signals of amplitudes +1V and −1V correspond to the logic states 1 and 0, respectively. Next, the input information is encoded into the phase of the spin wave. The conversion of the voltage signal into the spin wave phase is accomplished by the microstrip antenna. Each microstrip generates a local magnetic field to excite a spin wave in a ferromagnetic film. Depending on the polarity of the input signal, the initial phase of each spin wave may have a relative phase differences of $\pi$. Phases of "0" and "$\pi$" may be used to represent two logic states 1 and 0. Being excited, spin waves propagate through the spin waveguide or spin wave bus. The data processing in the circuit is accomplished by manipulating the relative phases of the propagating spin waves. The final logic state is detected by the inductive voltage measurement by receiving microstrip. The sign of the inductive voltage (positive or negative) corresponds to the final logic state. Then, the voltage signal may be amplified by conventional MOSFET to provide the compatibility with the external circuits, after completion of "computation".

To illustrate the concept, in Fig.1(b) we have shown a prototype logic circuit with two inputs and one output. The core of the structure consists of a ferromagnetic film (NiFe, for example) deposited on insulating substrate (SOI, for example) by a sputtering technique. The film may be entirely polarized along with the X axis when a magnetic field is applied. The thickness of the ferromagnetic layer is on the order of tens of nanometers. There are three asymmetric coplanar (ACPS) transmission strip lines on the top of the structure. These transmission lines are isolated from the ferromagnetic layer by the silicon oxide layer, and are used for spin wave excitation and detection. A voltage pulse applied to an input ACPS line produces a magnetic field perpendicular to the polarization of the ferromagnetic film, and, thus, generates spin waves, and the readout can be done by the output ACPS line. The structure is similar to one used for the time-resolved measurements of propagating spin waves [12]. In Fig.1(c) we show the equivalent circuit for the proposed logic circuit. The input and output circuits consist of the spin-based devices are depicted as LCR oscillators $(L_1, C_1, R_1)$, and the LCR transmission line $(L_0, C_0, R_0)$ is represented as the Spin Wave Bus. The oscillators are inductively coupled via the interaction with the ferromagnetic film. The film serves as a "magnetic" waveguide transmitting a magnetic field perturbation from one oscillator to another. The change of the current in any of the oscillators produces an inductive voltage in the others, and vice versa. Depending on the relative phase of the currents in the LCR circuits shown in Fig.1(c), the resultant inductive voltage at the central circuit can be maximal (in phase) or minimal (out of phase).

An elementary logic gates such as NOT, AND and OR can be realized on the prototype logic circuit shown in Fig.1(b). For example, the edge ACPS lines can be considered as the input ports, and the middle ACPS as the output port. The middle ACPS line detects the inductive voltage produced by the *superposition* of two waves. Depending on the relative phase of the spin waves, the amplitude of the inductive voltage may be enhanced (with two waves in phase) or decreased (with two waves out of phase) in comparison to the inductive voltage produced by a single spin wave. The relative phase is defined by the location of the ACPS lines and the polarity of the input excitation voltage signal. It is also possible to control the initial phase by adjusting the time of excitation, with these elementary sets is possible to realize different logic gates AND, OR, and NOT by controlling the *relative phases* of the spin waves. The detailed numerical simulations show the inductive voltage as a function of the relative phase of two spin waves as given in Ref.[10].

III. Spin Waveguide Materials

Thin film ferromagnetic material is the key element for the proposed spin wave bus circuits. First of all, room temperature operation (high Curie Temperature), long spin relaxation time are important challenges to be overcome. Second, for integration with Si technology, both materials and processing issues need to be addressed. Spin based devices require both soft and hard metals to produce desired functionality, and the materials are used in the recording (reading) magnetic hard discs. Hard ferromagnetic materials are used to pin (permanent) magnetic orientation, while soft ferromagnetic materials give the ability to change the spin polarization by an external magnetic or electric field. By properly choosing the materials, it is possible to fabricate logic circuits with different logic output. The best polarization obtained for elemental metals has been with Ni and Co, where up to 50% polarization has been obtained [13-15]. Going beyond elemental metals, alloys such as NiFe have been used. Among these alloys, the most promising metal alloy today for spintronic devices with the highest magnetic moment is $Fe_{0.65}Co_{0.35}$ [16].

Recently, in order to achieve the integration with Si electronics and to provide easy electric field control for next generation of Spintronics, great deal of efforts has been devoted to Diluted Magnetic Semiconductors (DMS). GaN and ZnO have been theoretically shown to possess Curie temperature above RT, that is Mn doped [17-19].

For (Ga, Mn)As, experimental results have shown a spin polarization of 85% at 110K [20]. Nevertheless, II-VI, III-V and group IV are still actively investigated. For all these materials low solubility of Mn and homogeneity are critical limiting factors. Similarly, oxide semiconductors such as Mn-ZnO [21] and Co-ZnO [22, 23] have also shown promise, and Mn solubility up to 10 mol% at room temperature Tc [24-26]. A review of literature [26] shows skepticism of ferromagnetism in DMS and oxide semiconductors as ferromagnetism believe to be induced by insoluble Co, Mn, Ni clusters. In addition, results vary and inconsistent polarization and Curie temperatures are reported [27].

More promising, are half metals from the perspective of Curie temperatures and spin polarization. Early theoretical studies indicated that NiMnSb [28] and $CrO_2$ [29] possessing a complete spin spilt band at the Fermi Level, predicting a polarization of 100%. These predictions have been experimentally confirmed with results coming close to the theoretical maximum [30]. Other half metals include $Fe_3O_2$, Heusler Metals (such as LSMO[31]), pervoskite (such as $(La_{1-x}A_x)MnO_3$ & $Sr_2FeMoO_6$). However, $CrO_2$ is by far the most researched half metal [30]. A variety of different anti- and ferromagnetic films on silicon have been grown and exchange integrals of composite materials have also been tabulated [32-35]. Figure 2 shows a summary of popular ferromagnetic materials gaining attention for potential use as spin waveguides.

IV. Signal propagation in RLC line and in Spin Wave Bus

Signal attenuation and signal group velocity are important parameters to be benchmarked for the spin wave bus. For these parameters, the cross-section dimensions and frequency operation range (from 1GHz to 100GHz) of the spin wave bus are the determining factors. In this section we estimate the transport characteristics of the spin transmission line and compare those parameters with these of the conventional electronic transmission line of the same dimension; we give the dimensions in Fig.3.

In Fig.3(a) we have schematically shown the general and the cross-sectional view of a microstrip line. The line consists of a conductive substrate (ground plane), a dielectric layer and a signal conductor on the top. The dielectric layer has thickness $t$ and relative permittivity $\varepsilon_r$, and the signal conductor has thickness $d$. The signal conductor width is $w$, and the ground conductor width is g. In Fig.3(b) we have shown the general and the cross-sectional view of the Spin Wave Bus. The bus consists of a ferromagnetic wire, which has the same dimensions as the signal conductor. The ferromagnetic film is shown on the top of non-magnetic insulating substrate. Signal propagation in both cases can be represented as a superposition of two waves traveling in the opposite directions along with the copper wire or the ferromagnetic film. The amplitude of the signal can be expressed as follows:

$$A(z,t) = A_1 e^{-\kappa z} \cos(\omega t - \beta z) + A_2 e^{\kappa z} \cos(\omega t + \beta z) \tag{1}$$

where $\kappa$ - represents the attenuation, and $\beta$ - defines the signal velocity $v = \omega/\beta$. These transport parameters let us to compare the performance of two transmission lines regardless the physical mechanisms of signal propagation.

First, we estimate signal losses in microstrip transmission line using a *RLC* model [36]. Here we use the formula for the transport coefficients given by:

$$\kappa = \frac{R}{2}\sqrt{2C/L}\left[1+\left(1+\frac{\omega_c^2}{\omega^2}\right)^{1/2}\right]^{-1/2}$$

$$\beta = \omega\sqrt{LC/2}\left[1+\left(1+\frac{\omega_c^2}{\omega^2}\right)^{1/2}\right]^{1/2}, \tag{2}$$

where $\omega_c = R/L$, $R$ is the resistance per unit length, $L$ is the inductance per unit length, and $C$ is the capacitance per unit length. At high-frequencies the resistance and inductance are subjects of significant modification due to the skin-effect. In order to take the skin-effect into consideration, we use the closed-form formulas found in Ref[37]. Here we reproduce the formula for high-frequency resistance and inductance per unit length [37]:

$$R'(f) = R'_0 + \frac{R'_\infty(f_s)\dfrac{\sqrt{f/f_s}+\sqrt{1+(f/f_s)^2}}{1+\sqrt{f/f_s}} - [R'_\infty(f_s) - R'_0]F(f) - R'_0}{1+\dfrac{k_r}{1+\omega/h}\log(1+f_s/f)}$$

$$L'(f) = \frac{R'_\infty(f_s)\sqrt{f/f_s}}{2\pi f(1+\sqrt{f_f/f})} + L'_\infty + \left[L'_0 - L'_\infty - \frac{R'_\infty(f_s)}{2\pi f_s}\right]F(f) ,$$

$$F(f) = \left[1+(f/f_0)\right]^{-1/2}, \quad f_0 = \frac{2}{\mu_0}\frac{R_w R_g}{R_w + R_g}, \quad f_s = \frac{k_s + \dfrac{10t/w}{1+w/d}}{\pi\mu_0\sigma t^2},$$

$$R'(0) = R_w + R_g, \quad R_w = \frac{1}{\sigma wt}, \quad R_g = \frac{1}{\sigma gt}, \tag{3}$$

where $R_w$ is the resistance per unit length of the signal conductor, $R_g$ is the resistance per unit length of the ground conductor, $k_s = 1.6$, $k_r = 0.2$ for microstrip lines. The resistance and inductance per unit length in the high-frequency region asymptotically behave as [37]:

$$R'(f) \to R'_\infty(f) = R'_\infty(f_i)\sqrt{f/f_i}$$

$$L'(f) \to L'_\infty + R'_\infty(f)/\omega \tag{4}$$

where $f_i$ is a chosen (reference) frequency, $R'_\infty$ denotes the skin-effect resistance per unit length, and $L'_\infty$ is the high-frequency external inductance given by $L'_\infty = \frac{\varepsilon_0 \mu_0}{C_0}$, where $\varepsilon_0 = 8.8542 \cdot 10^{-12}$ F/m is the permittivity of vacuum, $\mu_0 = 4\pi \cdot 10^{-7}$ H/m is the vacuum permeability, $C_0$ is the transmission line capacitance per unit length in vacuum. The reference skin-effect resistance and inductance were taken for a copper ($\sigma =56$ MS/m) microstrip line, w = 0.2 mm, d =0.1 mm, g =2mm, t =10 μm, at reference frequency 1 GHz: $R'_\infty$ =40.64 Ω/m, $L'_\infty$ =287.6 nH/m. The accuracy of the asymptotic formula for is estimated about 10% [37] . Then, we calculated the losses in the miscrostip line in the frequency range form 1GHz to 1THz. The results are shown in Fig.4. The signal attenuation is of the order of $10^3$ Np/m. The losses increases with frequency increase as with scaling down the thickness of the signal conductor. At high frequency ($\omega >> \omega_c$) the signal velocity approaches its limit – the speed of light in a given dielectric material.

The energy per bit in the metallic interconnect is limited by the by the Johnson noise

$$V_n = \sqrt{4k_B T Z B_L} \qquad (5)$$

where $k_B$ is the Boltzmann's constant, Z is the line impedance ($Z = \sqrt{\frac{\mu_0}{\varepsilon \varepsilon_0}} = 50\Omega$), the bandwidth of the transmission line $B_L$ must be larger than the inverse rise time $\tau_n$ ($B_L > [10\tau_n]^{-1}$), and the intrinsic switching time of a transistor is given by $\tau_n = \frac{CV_{DD}}{I_D}$.

Taking the gate length of 35nm, and $\tau_n = 2.5 ps$, we obtain that the noise floor is $V_n$=0.18 mV. Then for 6dB propagation loss and 3x noise margin, the minimum input voltage is $V_{in,met} = 2.2 mV$ and the energy per bit at 10GHz is $E_{met} = 10^{-18} J$. Even though this limit is proportional to $k_B T$, it is 3 orders of magnitude larger due to the factors in the noise limit.

Next, we describe the spin wave propagation in a ferromagnetic layer using the Landau-Lifshitz-Gilbert equation:

$$\frac{d\vec{m}}{dt} = -\frac{\gamma}{1+\alpha^2} \vec{m} \times \left[\vec{H}_{eff} + \alpha \vec{m} \times \vec{H}_{eff}\right], \tag{6}$$

where $\vec{m} = \vec{M}/M_s$ is the unit magnetization vector, $M_s$ is the saturation magnetization, $\gamma$ is the gyro-magnetic ratio, and $\alpha$ is the phenomenological Gilbert damping coefficient. The first term of equation (6) describes the precession of magnetization about the effective field and the second term describes its dissipation. $\vec{H}_{eff}$ is the effective field given as follows:

$$\vec{H}_{eff} = \vec{H}_d + \frac{2A}{M_s}\nabla^2 \vec{m} + \frac{2K}{M_s}(\vec{m} \cdot \vec{c})\vec{c} + \vec{H}_{ext}, \tag{7}$$

where $\vec{H}_d = -\nabla\Phi$, $\nabla^2\Phi = 4\pi M_s \nabla \cdot \vec{m}$, $A$ is the exchange constant, $K$ is the uniaxial anisotropy constant, and $\vec{c}$ is the unit vector along with the uniaxial direction, $\vec{H}_{ext}$ is the external magnetic field. We restrict our consideration by the magnetostatic waves propagating orthogonally to the film magnetization. In this case we use the analytical formula for spin wave dispersion in a finite-size ferromagnetic film [38] given by

$$\omega = \left[\omega_H(\omega_H + \omega_M) + \frac{\omega_M^2}{4}(1 - \exp^{-2kd})\right]^{1/2}, \tag{8}$$

where $d$ is the thickness of the ferromagnetic film, $\omega_H = \gamma H_0$, $\omega_H = \gamma 4\pi M_s$ $\gamma$ is the gyromagnetic ratio, $H_0$ is the external magnetic field at which film magnetization saturates at $M_s$. The attenuation of the spin waves is due to the magnon-electron, magnon-phonon and magnon-magnon scattering processes. According to the Landau-Lifshitz-Gilbert formalism (Eq.6), the combined effect of all scattering processes is described by the Gilbert damping coefficient $\alpha$. The decay time $\tau$ is given by

$$\tau = (2\pi\gamma\alpha M_s)^{-1}. \tag{9}$$

We estimate the spin wave attenuation $\kappa = (\tau v)^{-1}$ using Eqs. (8) and (9) for 100 nm thick NiFe film. In Fig.5 we show the results of numerical simulations showing spin wave attenuation as a function of frequency in the frequency range from 1GHz to 100GHz. In our numerical simulations we used experimentally found Gilbert coefficient $\alpha=0.0097$ from Ref. [12], and the following material characteristics for NiFe: $\gamma=19.91\times10^6$ rad/s Oe, $4\pi M_s$=10kG, $H_0$ =200Oe taken from the literature [39, 40].

Our estimates show the attenuation of the order of $10^3$ Np/m for GHz region with rapid increase up to $10^7$ Np/m in the high frequency region 100GHz. The cause of the losses increase is in the drop of spin wave phase velocity at high frequency. In the insert to Fig.5, we depict the phase velocity of the spin wave as a function of wave frequency. The velocity has order of $10^4$m/s at GHz region and exponentially decreases as the frequency increases. The results shown in Fig.5 are obtained for specific ferromagnetic material and specific spin wave propagation mode $k\perp H$ (magnetostatic surface spin wave). The phase velocity varies for different ferromagnetic materials and may be increased by applying an external magnetic field or/and modifying the size of the ferromagnetic wire. The fundamental limitation of the spin wave velocity is due to the finite strength of the exchange interaction between the neighbor spins in the lattice. The classical dispersion relation for the spin waves in bulk ferromagnets in the long wavelength limit $ka<<1$ ($a$ is the lattice constant) can be rewritten in the following form:

$v_p \approx \sqrt{\dfrac{2\omega J s a^2}{\hbar}}$, where $J$ is the exchange energy. The higher is the exchange energy the faster are the spin waves in a given ferromagnetic material. For all experimentally studied ferromagnetic materials having high Curie temperature (for example, cobalt and iron having 1388 K and 1043 K Curie temperature, respectively,) the spin wave phase velocity does not exceed $10^5$ m/s. Thus, this velocity can be taken as a benchmark for the maximum signal phase velocity in spin wave bus.

V. Spin wave energy

Spin wave energy $E_{sw}$ can be estimated on very general thermodynamic grounds as follows:

$$E_{sw} = \mu_0 \Delta M H_{ext} V, \qquad (10)$$

where $\Delta M$ is the magnetization change (spin wave amplitude) caused by the external magnetic field of the strength $H_{ext}$, and $V$ is the material volume. The equation is similar to one presented in Ref. [41] for magnetic domain-wall logic, except the magnetization change produced by a spin wave is much less than the saturation magnetization $M_s$, $\Delta M \ll M_s$. In turn, the amplitude of the external magnetic filed required to excite a spin wave is much less than that is required to reverse the magnetization of the whole domain. It is important to recognize a propagating domain-wall and a spin wave as two different spin transport mechanisms. Spin wave can be considered a small perturbation of the domain magnetization, and the amplitude of the spin wave can be scaled down with no limit to $M_s$. The energy of a spin wave may be close to $k_B T$, regardless the material volume. At the same time, the total energy required to excite the spin wave depends on the efficiency of a given excitation mechanism. As an example, we would like to refer to the spin wave excitation by local magnetic field generated by microstrips [42]. For a simple geometry (circular loop of radius *R*), and assuming the distance to the ferromagnetic film much less than the antenna radius $d<<R$, the strength of the magnetic field produced by antenna is given by:

$$H_{ext} \approx \frac{I_{ext}}{2\pi R}, \tag{11}$$

where $I_{ext}$ is the excitation current. The strength of the magnetic filed required for local magnetization change $\Delta M$, i.e., the amplitude of the spin wave, can be estimated from Eq.(10) as follows:

$$\frac{\Delta M}{M_s} \approx \gamma \tau_{ext} H_{ext} \tag{12}$$

where $\tau_{ext}$ is the duration of the excitation pulse. Using Eq.(11) and (12), we estimate the energy required for spin wave excitation $E_{ext}$ as follows:

$$E_{ext} = I_{ext}^2 \cdot Z \cdot \tau_{ext} = \left(\frac{\Delta M}{M_s}\right)^2 \left(\frac{2\pi R}{\gamma}\right)^2 \frac{Z}{\tau_{ext}} \tag{13}$$

where $Z$ is the impedance. Taking $\Delta M / M_s = 0.01$, $R = 2\mu m$, $Z=50\Omega$, $\gamma=19.91\times10^6$ rad/s/Oe ($4.0\times10^4$ m/As) from [12], and $\tau_{ext} = 100$ps, we have $E_{ext} = 4.9\times10^{-12}$ J. For comparison, the spin wave energy obtained for the same parameters using Eq.(10) is $E_{sw}=6.4\times10^{-17}$ J. Only a small part of the excitation energy will be converted into the spin wave energy $E_{sw} << E_{ext}$. The excitation efficiency can be enhanced by optimizing the antenna structure and prolonging the time of excitation. At any rate, the use of microstrip antenna is the most convenient but not the optimum way for spin wave excitation.

The most important question is: "How small energy spin wave can be detected by the tools of the conventional technology?" There are several techniques known for spin wave detection: neutron scattering [43], optical measurements [44], magnetoresistance measurements [45] and inductive voltage measurement [40]. We consider the time resolved inductive voltage measurements as the most convenient technique for spin wave detection. According to the Faraday's law, the magnitude of the inductive voltage is proportional to the speed of the magnetic flux change $V_{ind} = -d\Phi/dt$, where $\Phi$ is the magnetic flux through the area $S$ of the detecting conducting contour. The sign of the inductive voltage is defined by the rate of the flux change. Thus, two spin waves having $\pi$ phase difference can be recognized by the sign of the produced inductive voltage. An inductive voltage signal of the order of several mV produced by spin waves propagating through a 27nm thin ferromagnetic film ($Ni_{81}Fe_{19}$) was clearly detected at room temperature [12]. The magnitude of the inductive voltage is proportional to the amplitude of the spin wave, contour area and the spin wave frequency. The peak voltage can be estimated as follows:

$$V_{max} \approx \mu_0 \Delta M S \omega. \qquad (14)$$

Using Eqs. (10) and (14), we plotted in Fig.6 the maximum amplitude of the inductive voltage versus the energy of the spin wave in NiFe 100nm thick film. The amplitude of the inductive voltage can be enhanced by increasing the effective area of the detecting contour (multiple conducting loops), and increasing the spin wave frequency. The spin wave signal has a specific signature – it provides an oscillating output of a certain frequency $V_{in} \approx V_0 \sin(\omega t)$. By applying a references signal $V_{ref} \approx V_1 \sin(\omega t + \theta)$, it is possible to extract the harmonic component in the form of a DC output $V_{DC} = 0.5\sqrt{V_0 V_{ref}} \cos(\theta)$. The current technique, using standard phase lock-in amplifiers allows us to detect $nV$ (at room temperature). This fact is in favor of spin waves showing the possibility to minimize the operation voltage and decrease power consumption. The use of lock-in amplifier requires additional energy consumption, and can be implemented efficiently only for final read-out operation.

Magnetoresistance measurements may be an alternative technique for spin wave detection. It was experimentally found that spin waves play a prominent role in magnetic tunneling junction conductance [45]. The change of the conductance due to the spin waves can be expresses as follows:

$$\Delta \rho_{mag} \propto \frac{BT}{D(T)^2} \ln\left(\frac{\mu_B B}{k_B T}\right) \tag{15}$$

where $B$ is the inductance, $\mu_B$ is the Bohr magneton, and $D(T)$ is the temperature dependent constant $D(T) = D_0 - D_1 T^2 - D_2 T^{5/2}$, ($D_0 \sim 10^2$, $D_1 \sim 10^{-6}$, $D_2 \sim 10^{-8}$ meVA$^2$ at room temperature for Ni film [45]). However, it is not clear whether or not the magnetoresistance measurements can be used for spin wave phase recognition.

IV Discussions

There are many potential advantages for logic circuits with Spin Wave Bus. (i) The local interconnects problem may be resolved, as there are no conducting wires for local interconnection. All communications between the input/output terminals and Spin Wave

Bus are via magnetic field. The coupling occurs in a wireless manner by a magnetic field produced by spin waves. (ii) Enhanced logic functionality. The utilization of wave superposition offers the possibility to accomplish useful information processing inside the bus, without the use of additional logic elements. A number of logic gates can be realized in one logic circuit. (iii) Parallel data processing. A number of bits can be encoded in spin waves of different frequencies and processed at the constant time. (iv) Defect tolerance. A new phenomenon of momentum relaxation reversal has been discovered experimentally and explained theoretically for spin waves in ferromagnetic films [46]. It is shown that the process of momentum relaxation, caused by the scattering of a signal wave on defects, can be reversed, and the signal can be self-reconstructed after it left the scattering region. The reversal of momentum relaxation is achieved by frequency selective parametric amplification of a narrow band of scattered waves having low group velocities and frequencies close to the frequency of the original signal wave. The effect of momentum relaxation reversal can be used for signal self-correction. (v) Wireless read-in and read-out procedures. By using the micro- or nano-scale antennas it is possible to excite and detect spin waves in a wireless manner. (vi) CMOS compatible processing. There is no major departure for CMOS fabrication. The layered structure of the device can be easily realized with high accuracy by film deposition by sputtering technique.

A fundamental drawback inherent to the spin wave bus is relatively low spin wave velocity ~ $10^5$ m/s [12]. The low phase velocity inevitably results in significant time delay for signal propagation. It can be compensated, in part, by taking into account short (of the order of hundred nanometers) distances among the spin-based devices. Another significant drawback is high signal attenuation. The fundamental cause of spin wave amplitude damping is the scattering on phonons. It has been shown experimentally and verified theoretically that magnetic dissipation plays a significant role in propagation of spin waves. Due to the dissipation even in the high quality low-loss yttrium iron garnet (YIG) films the propagation length has been found not to exceed one centimeter. The typical spin wave life-time is limited to the order of a few hundred nanoseconds [47]. We would like to outline, that the utilization of spin wave for information transmission may be applicable only for short-range in-chip interconnects which could possibly find suitable use in cellular array based architectures.

Another important issue is the possibility of obtaining signal gain in spin wave bus. There are two possible approaches to this problem: (i) amplify spin wave during propagation in the waveguide, and (ii) amplify the magnetization change produced by the spin wave at the recipient device. To amplify a propagating spin wave perpendicular and parallel microwave pumping can be applied [48]. A possibility to compensate spin wave's damping and to amplify spin waves propagating in thin ferromagnetic films by parallel pumping has been proven experimentally [49, 50]. A parametric microwave spin wave amplifier giving the gain coefficients up to 40 dB for the input power levels about 1 pW has been demonstrated experimentally [49]. The use of the microwave pumping may not be efficient from a power consumption point of view, as it inevitably results in heat dissipation inside the spin wave bus. It would be more desirable to keep the energy for device-to-device communication minimal and amplifying only the final output signal. In other words, it is desirable to amplify the magnetization change caused by the spin wave at the recipient device rather then pump the spin waves in the bus. A possible solution is the combining of spin wave bus with a diluted magnetic semiconductor (DMS) structure and use the effect of hole-mediated ferromagnetism for signal amplification. In this case, the gain is a function of the DMS cluster size and the critical hole concentration required for the paramagnetic-to-ferromagnetic transition. By estimate, the amplification might be as high as $10^5$-$10^6$. at the cost $10^{-17}$ J for DMS volume 20nm×20nm×100nm, at the critical hole density $10^{20}$ cm$^{-3}$ [51].

Cross-talks and capacitive coupling among the transmission lines may be significant with scaling down the separation distance between the excitation/detection ports. The problem of input-output isolation and signal unidirectional propagation is one of the critical design issues for the proposed spin wave based logic devices and deserves special consideration.

V Conclusions

We presented a feasibility study of logic circuits with spin wave bus. The use of spin waves offers an original way to build a wireless nanoscale architecture, where information transmission and processing can be accomplished without a charge transfer. The interaction between Spin Wave Bus and the other devices is via a magnetic field produced by spin waves. High losses ($10^6$ Np/m) restrict the potential use of spin waves for information transmission to micrometer range only. Signal propagation in Spin Wave Bus is limited due to the low group velocity ($10^5$ m/s at most). Energy per bit in the Spin Wave Bus can be scaled down to *1000 $k_BT$* at room temperature limited by the noise factor. Although the efficiency of spin wave excitation by microstrip antenna is very low, and the energy required for spin wave excitation exceed the energy of the spin in several orders of magnitude. Signal detection can be accomplished by (i) inductive voltage measurements and (ii) magnetoresistance measurements. Potentially, the use of spin wave may be the most efficient way for spin-based devices interconnection. A set of logic gates can be realized in one module consisting of a number of devices united with Spin Wave Bus. Finally, we would like to summarize our conclusions on Spin Wave Bus:

- Spin wave bus is inferior to traditional metal interconnects in all figures of merit;
- Low group velocity limits the signal propagation time and causes significant delay inherent for spin wave bus;
- Spin wave excitation by the electric currents is energetically inefficient;
- Logic devices with spin wave bus requiring a magnetization-to-voltage conversion in each logic element are inferior too;
- The value of spin wave bus is an interface between electronic circuits and integrated spintronics circuit;
- The logic circuits with spin wave bus allow us to provide wireless read-in and read-out.

Acknowledgement

We would like to thank Dr. G. I Bourianoff for the insightful discussion. The work was supported in part by the MARCO-FENA center and by the Western Institute of Nanoelectronics.

**Figure captions**

Fig.1 (a) The general concept of logic circuits with SWB. The operation procedure includes: input signal conversion into the phase of spin wave; data processing using the effect of spin wave superposition; read-out by inductive voltage measurements. (b) An example of a logic circuit with SWB. Three ACPS transmission lines are made on top of a ferromagnetic layer. The lines and the ferromagnetic layer are isolated by the oxide layer. Each of the ACPS lines can be used for spin wave excitation and detection. Spin waves propagate through the ferromagnetic film referring as "Spin Wave Bus" and can be detected by the time-resolved inductive voltage measurements. (c) The equivalent circuit consists of the inductively coupled oscillators. The LCR transmission line stands for the Spin Wave Bus.

Fig.2   Summary of ferromagnetic materials for potential use as spin waveguides.

Fig.3 (a) General and cross-sectional view of the microstrip line. From the bottom to the top: conductive substrate of thickness *d*; dielectric layer of thickness *t* and relative permittivity $\varepsilon_r$; signal conductor of thickness *d*. The signal conductor width is *w*, and the ground conductor width is g. (b) General and cross-sectional view of the spin wave bus. A ferromagnetic wire is on the top of a non-magnetic insulating substrate. The wire is polarized in the z direction. Wire width is w and thickness *d*.

Fig.4 Results of numerical simulations on signal propagation in microstrip line taking into account skin-effect. Signal attenuation as a function of frequency for different size of the signal conductor 10μm, 1μm, and 0.1μm.

Fig.5 Results of numerical simulations on signal propagation in NiFe Spin Wave Bus. Signal attenuation as a function of frequency for 0.1μm thick film. The insert depicts the signal phase velocity as a function of frequency.

Fig. 6 Inductive voltage amplitude as a function of the spin wave energy. Numerical simulations for 100nm NiFe wire. The lines A,B, and C correspond to the spin wave frequencies 1GHz, 10GHz, and 100GHz, respectively.

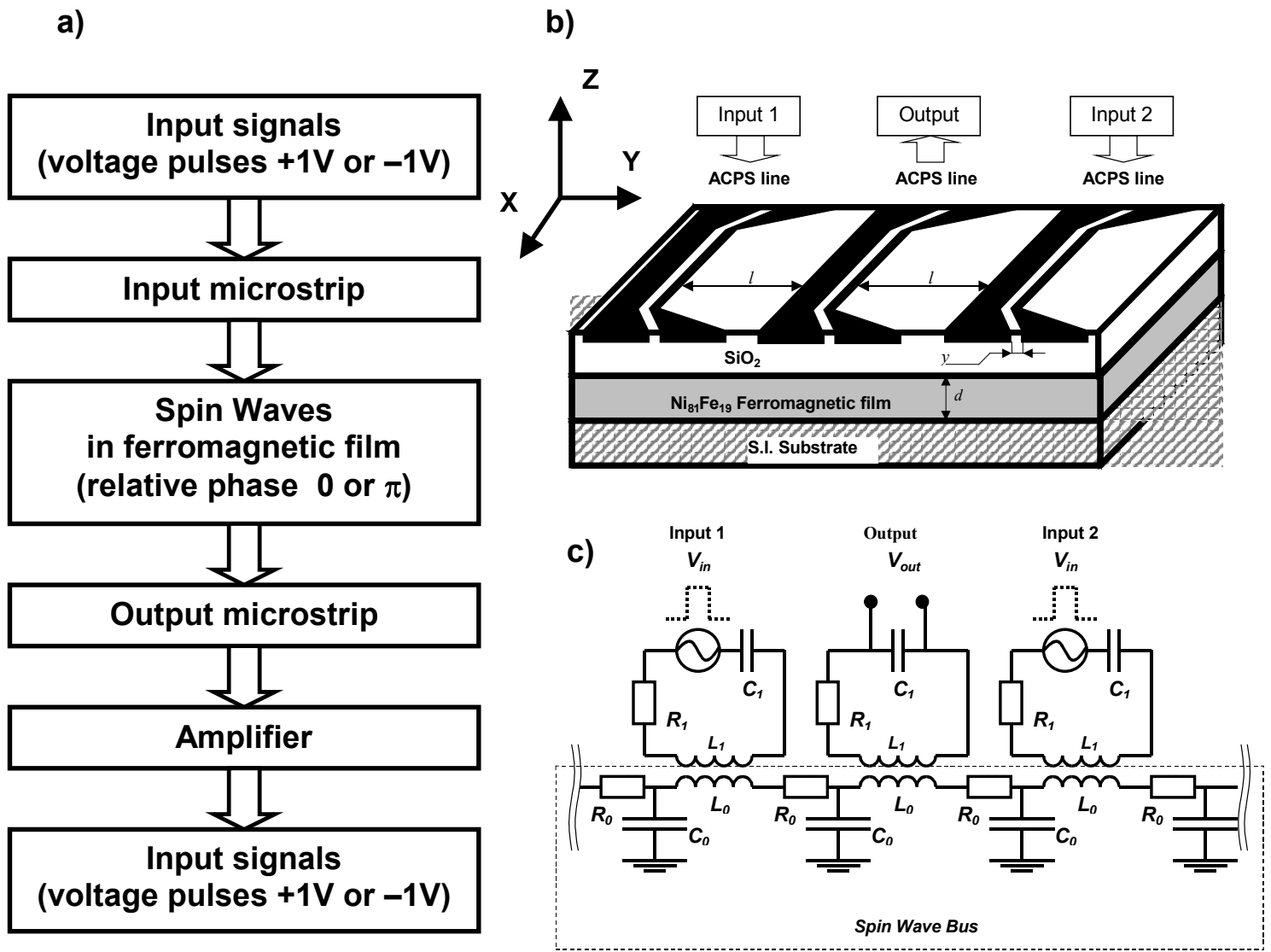

Fig.1

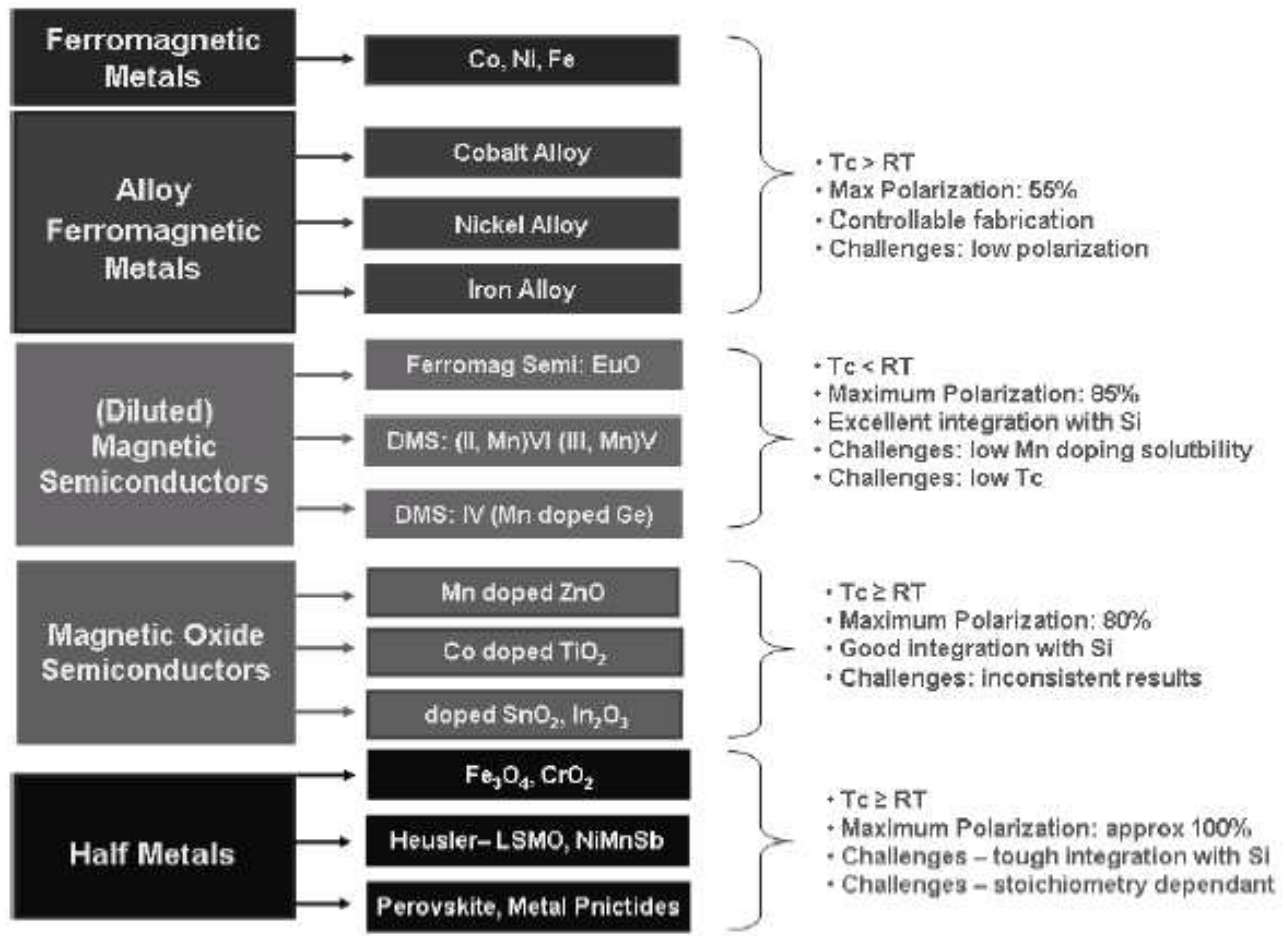

Fig.2

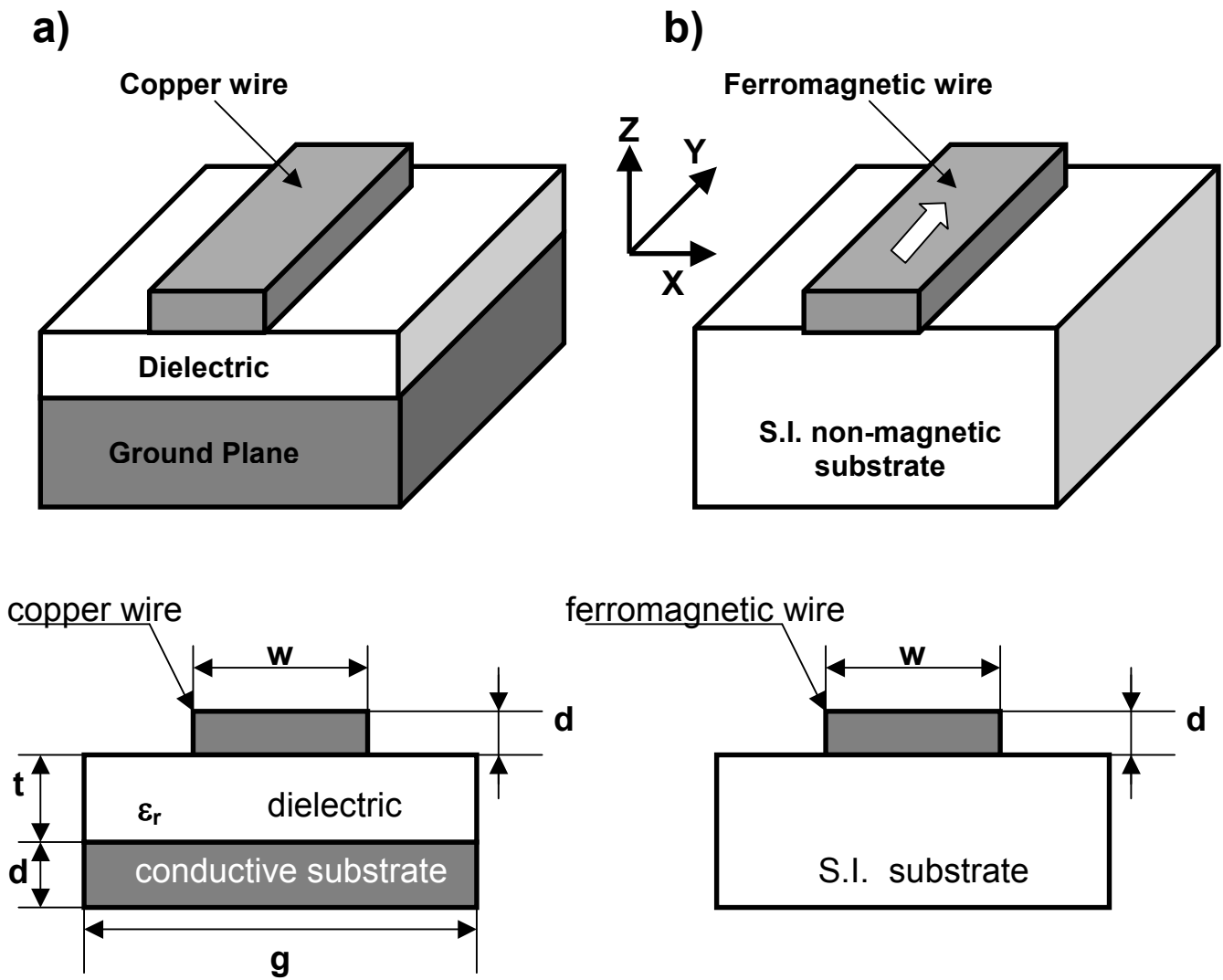

Fig.3

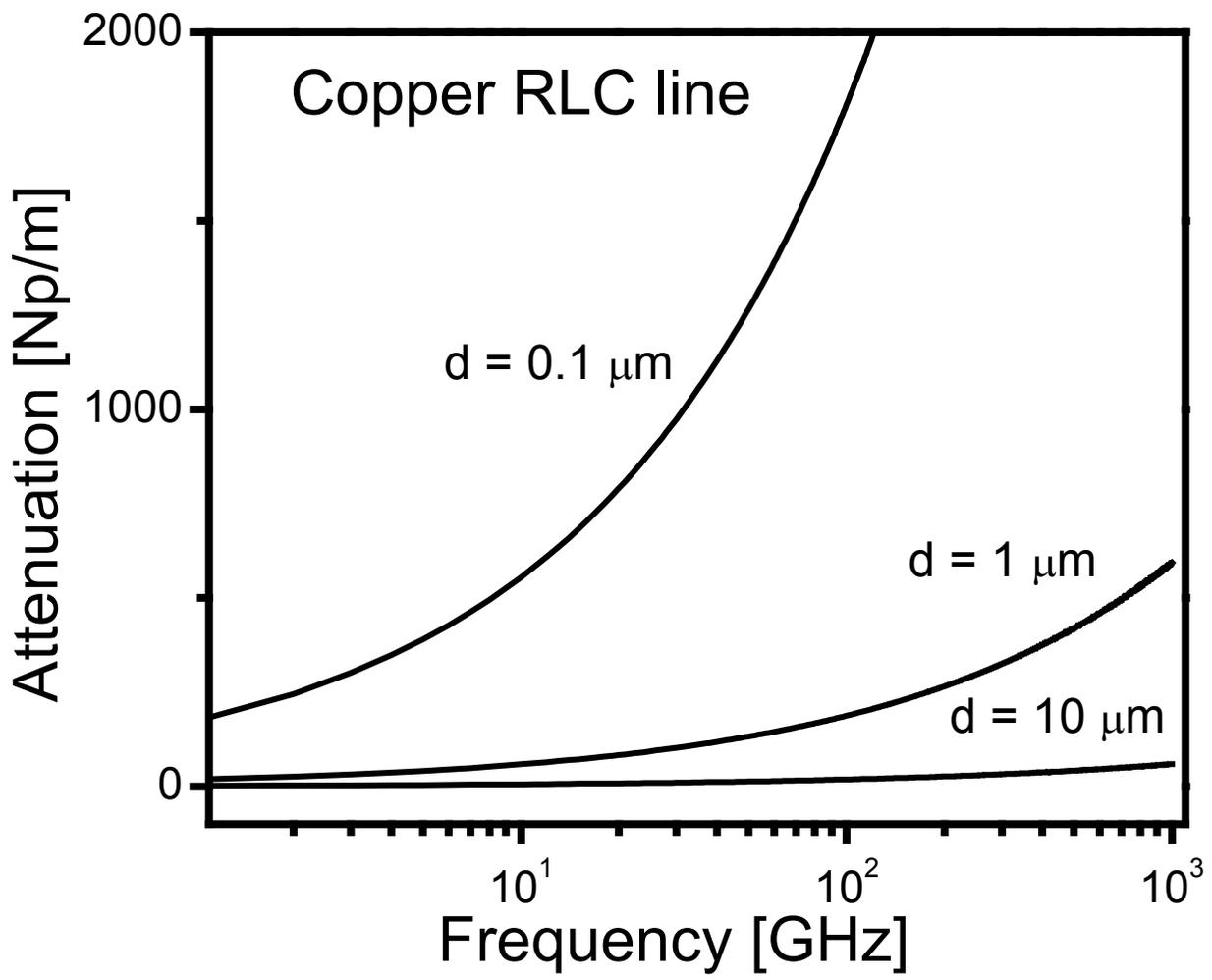

Fig.4

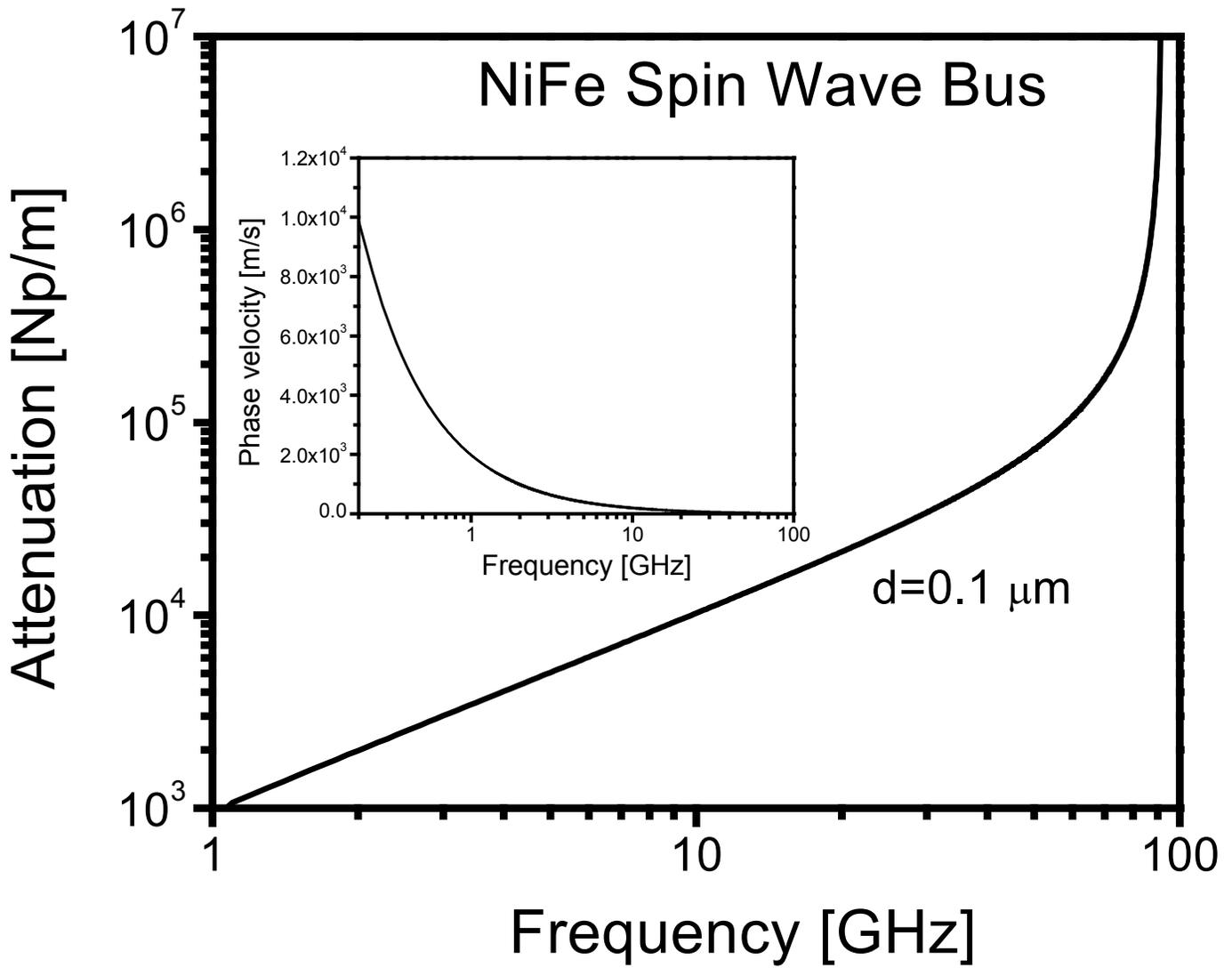

Fig.5

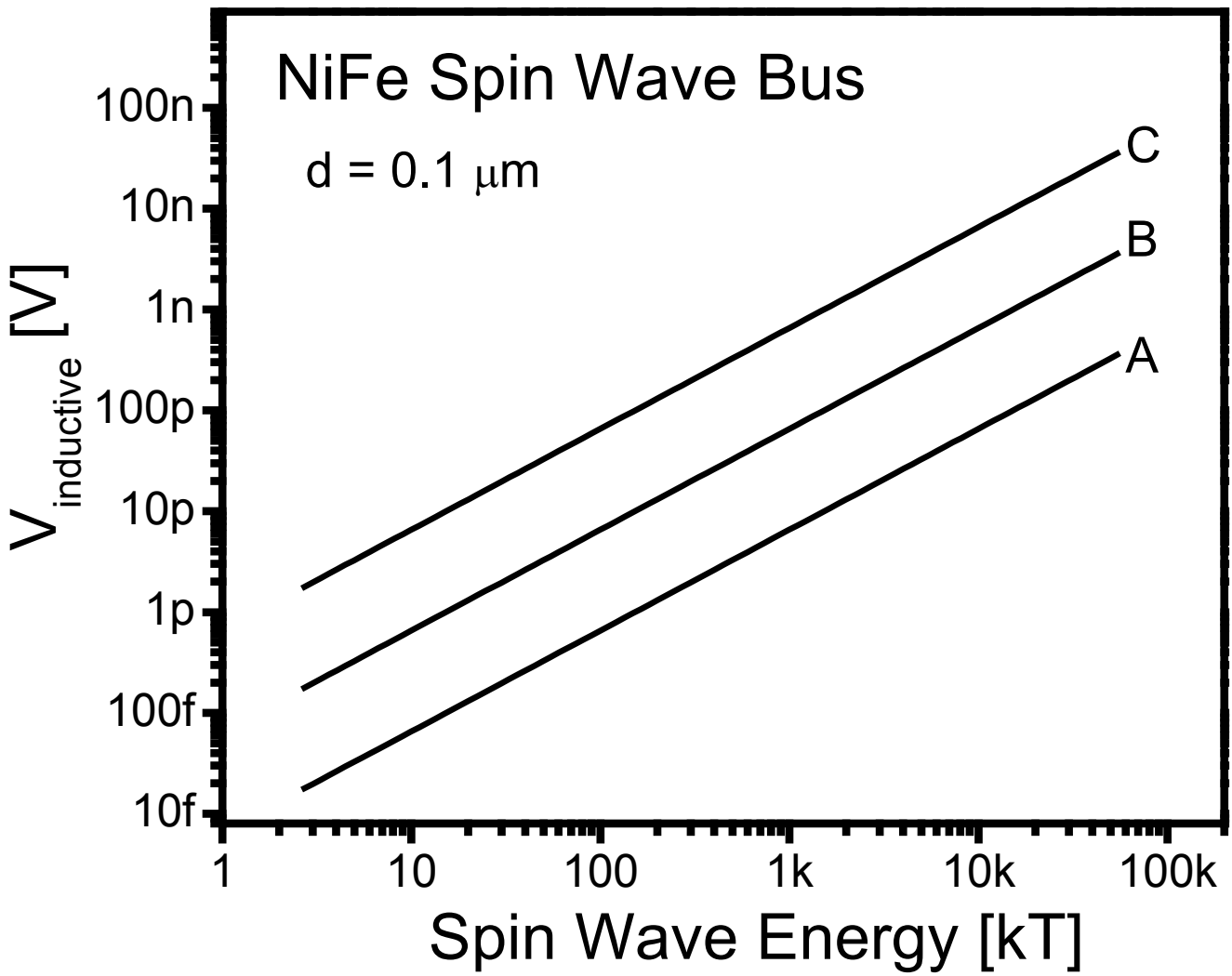

Fig.6